\documentclass[fleqn,usenatbib]{mnras}

\usepackage{newtxtext,newtxmath}

\usepackage[T1]{fontenc}

\DeclareRobustCommand{\VAN}[3]{#2}
\let\VANthebibliography\thebibliography
\def\thebibliography{\DeclareRobustCommand{\VAN}[3]{##3}\VANthebibliography}

\usepackage{graphicx}	
\usepackage{amsmath}	
\usepackage{multirow}

\title[Identification of a new ULX in NGC 1316]{Identification of a New Ultraluminous X-ray Source in NGC 1316}

\author[S. Allak et al.]{S. Allak,$^{1,2}$\thanks{E-mail:0417allaksinan@gmail.com} 
A. Akyuz,$^{1,3}$
N. Aksaker,$^{1,4}$
M. Ozdogan Ela,$^{1}$
S. Avdan,$^{1}$
F. Soydugan$^{2,5}$
\\
$^1$Space Science and Solar Energy Research and Application Center (UZAYMER), University of Çukurova, 01330, Adana, Turkey\\
$^2$Department of Physics, University of Çanakkale Onsekiz Mart, 17100, Çanakkale, Turkey\\
$^3$Department of Physics, University of Çukurova, 01330, Adana, Turkey\\
$^4$Adana Organised Industrial Zones Vocational School of Technical Science, University of Çukurova, 01410, Adana, Turkey\\
$^5$Astrophysics Research Centre and Ulupınar Observatory, University of Çanakkale Onsekiz Mart, 17100, Çanakkale, Turkey\\
\\
}
\date{Accepted 2020 October 1. Received 2020 September 30; in original form 2020 July 20.}

\pubyear{2020}

\begin{document}
\label{firstpage}
\pagerange{\pageref{firstpage}--\pageref{lastpage}}
\maketitle

\begin{abstract}
In this study, we report identification of a new ultraluminous X-ray source (ULX) named as X-7 in NGC 1316, with an unabsorbed luminosity of 2.1$\times$10$^{39}$ erg s$^{-1}$ using the two recent {\it Chandra} archival observations. The X-7 was detected in the {\it Chandra} 2001 observation and was included in the source list of the NGC 1316 as CXOUJ032240.8-371224 with a luminosity of 5.7$\times$10$^{38}$ erg s$^{-1}$. Present luminosity implies a luminosity increase of a factor of $\sim$ 4. The best fit spectral model parameters indicate that X-7 has a relatively hot disk and hard spectra. If explained by a diskblackbody model, the mass of compact object is estimated as $\sim$ 8 M$\odot$ which is in the range of a stellar-mass black hole. The X-7 shows a relatively long-term count rate variability while no short-term variability is observed. We also identified a unique optical candidate within $0.22\arcsec$ error circle at 95\% confidence level for X-7 using the archival {\it HST}/ACS and {\it HST}/WFC3 data. Absolute magnitude (M$_{V}$) of this candidate is -7.8 mag. Its spectral energy distribution is adequately fitted a blackbody model with a temperature of 3100 K indicating an M type supergiant, assuming the donor star dominates the optical emission. In addition, we identified a transient ULX candidate (XT-1) located $6\arcsec$ away from X-7 has a (high) luminosity of $\sim$ 10$^{39}$ erg s$^{-1}$ with no visible optical candidate.
\end{abstract}



\section{Introduction}

Ultraluminous X-ray sources (ULXs) are described as off-nucleus objects with high X-ray luminosities in the range of 10$^{39}$ $-$ $10^{41}$ erg s$^{-1}$ in external galaxies (see the review by \citealp{2017ARA&A..55..303K}). For most, their high luminosities are thought to be produced by super-Eddington accretions onto stellar mass black holes or neutron stars \citep{2013MNRAS.432..506P, 2013MNRAS.435.1758S, 2013ApJ...778..163B, 2017Sci...355..817I}. For some, sub-Eddington accretions onto intermediate mass black holes (IMBH) are possible \citep{1999ApJ...519...89C, 2001ApJ...551L..27M, 2015MNRAS.448.1893M}. Actually, numerous studies have examined the nature of the ULX systems. ULX X-2 of M82 with its coherent pulsations was the first neutron star in such systems \citep{2014Natur.514..202B}. Since then, the number of known pulsating ultraluminous X-ray sources (PULXs) has risen to 6 \citep{2016ApJ...831L..14F, 2017Sci...355..817I, 2017MNRAS.466L..48I, 2018MNRAS.476L..45C, 2019MNRAS.488L..35S, 2019arXiv190604791R}. In addition, a cyclotron resonance scattering feature in the X-ray spectrum of the source ULX-8 in M51 was discovered by \cite{2019MNRAS.486....2M} assuring an important hint about the nature of the magnetic field in ULXs that hosting neutron stars.

High quality data from {\it XMM-Newton} and {\it Chandra} revealed that the spectra of ULXs are different from those observed in the Galactic black hole binaries (GBHBs). A notable feature in the spectra of most ULXs is an unambiguous curvature in its spectrum between 3 to 7 keV. This curvature can be produced through Comptonization of a cold, optically thick corona around the compact object or alternatively in the inner regions of a geometrically thick accretion disk \citep{2006MNRAS.368..397S,2007Ap&SS.311..203R,2009MNRAS.397.1836G}. Based on these spectral features, a new ultraluminous accretion state with super-Eddington accretion occurring onto a black hole or a neutron star has been proposed by \cite{2009MNRAS.397.1836G}. The extended high energy interval of {\it NuSTAR} observations of ULXs clarified that the curvature was extending above 10 keV. This feature is also different from GBHB, whose cutoff energy is over 60 keV \citep{2013ApJ...778..163B,2016MNRAS.460.4417L,2017xru..conf..314P,2017ApJ...834...77F}.

Determining the optical counterparts of ULXs plays an important role to understand the origin of emission (i.e., accretion disk and/or donor star) and to estimate the mass, age and spectral type of the donor. 
However, identification of the optical counterparts of ULXs is difficult due to large extragalactic distances to them, some of which exhibiting a disk irradiation in a binary system.
The observed optical candidates of ULXs are too faint (m$_{V}$ > 21 mag.) to study in detail from HST images and/or confirm spectroscopically from the ground-based observations \citep{2008A&A...486..151G, 2015MNRAS.452.1112E, 2015NatPh..11..551F,2017ARA&A..55..303K,2019ApJ...877...57Q}. However, there are several exceptions in which the type and mass of the donor star are determined by optical spectroscopy; for instance: ULX-1 of M101 has a Wolf-Rayet donor \citep{2013Natur.503..500L} while ULX-1 of NGC 253, ULX-2 of NGC 925 and ULX-1 of NGC 4136 \citep{ 2015MNRAS.453.3510H, 2016MNRAS.459..771H} have M-type supergiant donors and P13 of NGC 7793 it is a PULX has a blue supergiant donor of type B9Ia \citep{2011AN....332..367M,2014Natur.514..198M}.

In general, absolute magnitudes of ULXs, M$_{V}$, are in the range of $-$8 and $-$3 mag. \citep{2018ApJ...854..176V}. Although there are hundreds of ULXs known, only about 20 have singled out optical counterparts \citep{2011ApJ...737...81T,2013ApJS..206...14G,2019ApJ...875...68A}. For many other ULXs, multiple undecidable optical counterparts have been detected \citep{2005MNRAS.356...12S,2007ApJ...658..999M,2016ApJ...828..105A,2019ApJ...875...68A,2019MNRAS.488.5935A}. Mostly, they have very faint optical candidates showing star-like spectral energy distributions (SEDs). Their SEDs are constructed by assuming that the optical emissions come from donor stars \citep{2012ApJ...745..123G, 2016MNRAS.455L..91A,2018ApJ...854..176V}.

NGC 1316 (also known as Fornax A) is a giant elliptical galaxy located in the outskirts of the Fornax galaxy cluster. In many studies, the distance of NGC 1316 was estimated with slightly different values from each other \citep[e.g.][]{2003ApJ...583..712J,2007ApJ...657...76F,2010AJ....140.2036S,2013A&A...552A.106C,2018ApJ...866..145H,2018MNRAS.481.4472L,2019ApJ...887..149B}. Throughout this work, we adopted the distance to the NGC 1316 as 19 Mpc \citep{2003ApJ...583..712J}. In the radio band, NGC 1316 is one of the brightest galaxy in the sky, with giant double radio lobes and well defined core jet structure \citep{1984AJ.....89.1650G}. The galaxy has a peculiar morphology with numerous tidal tails, shells and loops of interstellar medium \citep{1980ApJ...237..303S, 1981ApJ...246..722S}. NGC 1316 also displays dust patches in the central region together with the prominent dust lanes oriented along its optical minor axis (see Fig. \ref{F:1}).

In the previous studies, it is reported that the galaxy hosts many X-ray sources including several ULXs by using the archival {\it Chandra} and {\it XMM-Newton} data \citep{2003ApJ...586..826K,2004ApJS..154..519S,2011ApJS..192...10L,2019MNRAS.483.5554E}. Based on {\it Chandra} observation of 2001, \citet{2003ApJ...586..826K} detected 81 point sources within the 25th magnitude isophotal ellipse of NGC 1316. According to their source list, they reported only one potential ULX source (\#29) even though three more sources (\#12, \#13 and \#44) have luminosities $\geq$ 10$^{39}$ erg s$^{-1}$. They also reported that the luminosity of the \#13 is highly uncertain due to the contamination of the X-ray diffuse emission and remaining two sources are located at the center of galaxy with elongated shapes and uneasy source definitions. On the other hand, \cite{2011ApJS..192...10L} analyzed the same data and determined 4 ULXs in NGC 1316. These ULXs were X-4 (\#29), X-6 (\#64), X-5 (\#13) and X-3 (\#16, beyond D$_{25}$\footnote{D$_{25}$; defined by the 25 mag arcsec$^{-2}$ brightness level.} ellipse of NGC 1316) with numbering by their maximum detection significance. Recently, \cite{2019MNRAS.483.5554E} have reported the aforementioned ULX candidates by using {\it XMM-Newton} data. 

In the present study, we report the identification of the X-ray source CXOUJ032240.8-371224 as a new ULX to be denoted as ULX X-7 (hereafter X-7) as a result of the analysis of 2019 {\it Chandra} data. We obtained the peak luminosity of X-7 as 2 $\times $10$^{39}$ erg s$^{-1}$ in the energy range of 0.3–8 keV at a distance of 19 Mpc. However, this source has earlier been cataloged as \#14 by \cite{2003ApJ...586..826K} with a luminosity of $\sim$ 5.75 $\times$ 10$^{38}$ erg s$^{-1}$ in the same energy range. We further investigated the X-ray spectral and temporal properties of X-7. We also identified a single optical counterpart of this source using the {\it HST} multi-band optical observations.

The present paper is organized by as follows: {\it Chandra} and {\it HST} data reductions and analysis are presented in Section \ref{section:2}. The results of the analysis and discussions are given in Section \ref{section:3}. 

\section{Observations, Data Reduction and Analysis} \label{section:2}

\subsection{X-rays} 

NGC 1316 was observed by {\it Chandra} ACIS-S once in 2001, and five times in 2019. In addition, this galaxy was observed four times by {\it XMM-Newton} between 2005 and 2009 but only two of these observations included the position of X-7. However, the spatial resolution of {\it XMM-Newton} is not high enough to resolve the position of X-7 in these two observations. There were numerous observations of NGC 1316 obtained by {\it Swift-XRT} between 2006 and 2020. Also the source X-7 could not be resolved clearly in these observations. Therefore, only the data taken with {\it Chandra} ACIS-S observations were used in the present study. The log of observations is given in Table \ref{T:1}.

\begin{figure*}
\begin{center}
\includegraphics[scale=0.41]{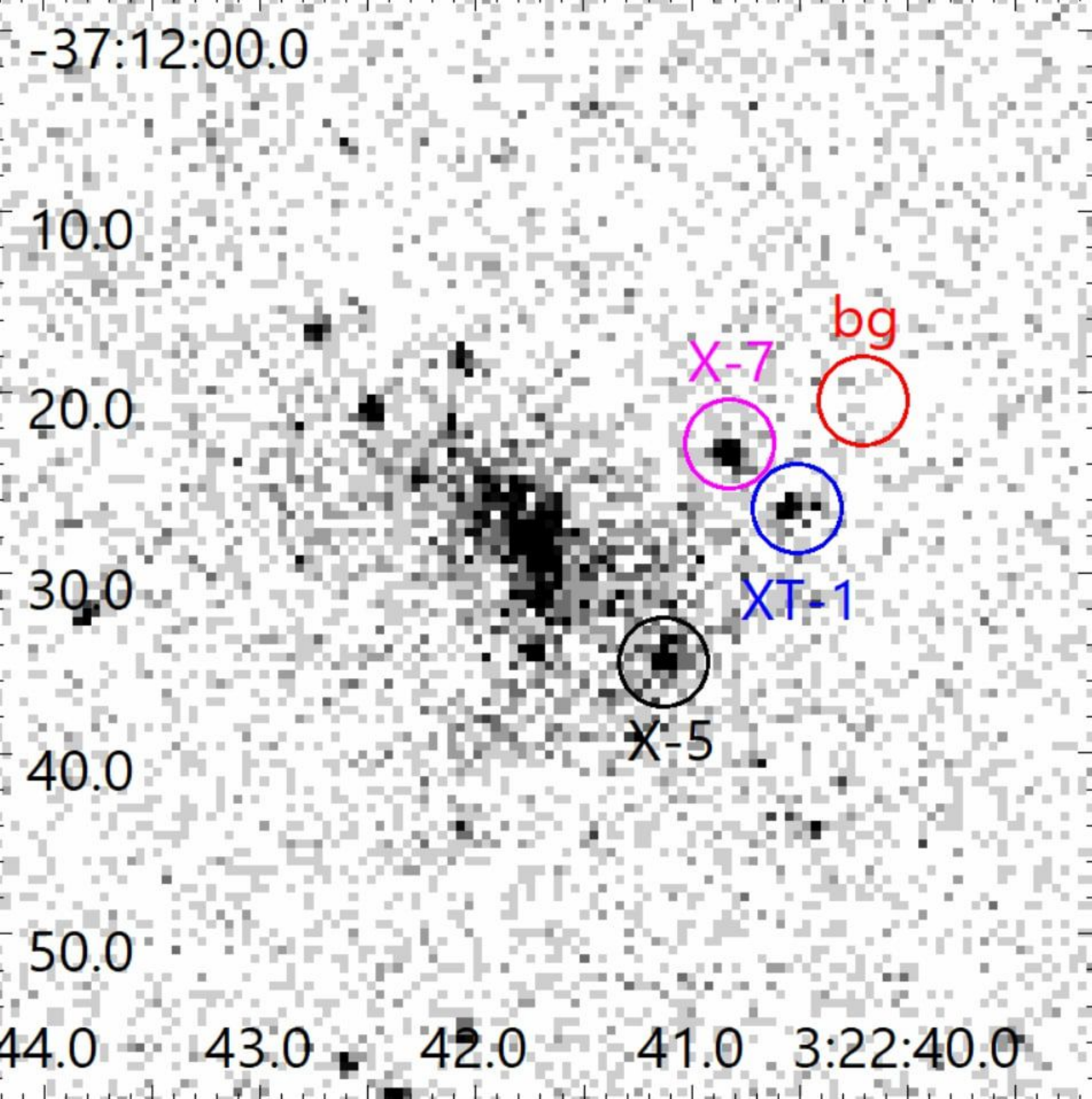}
\includegraphics[scale=0.41]{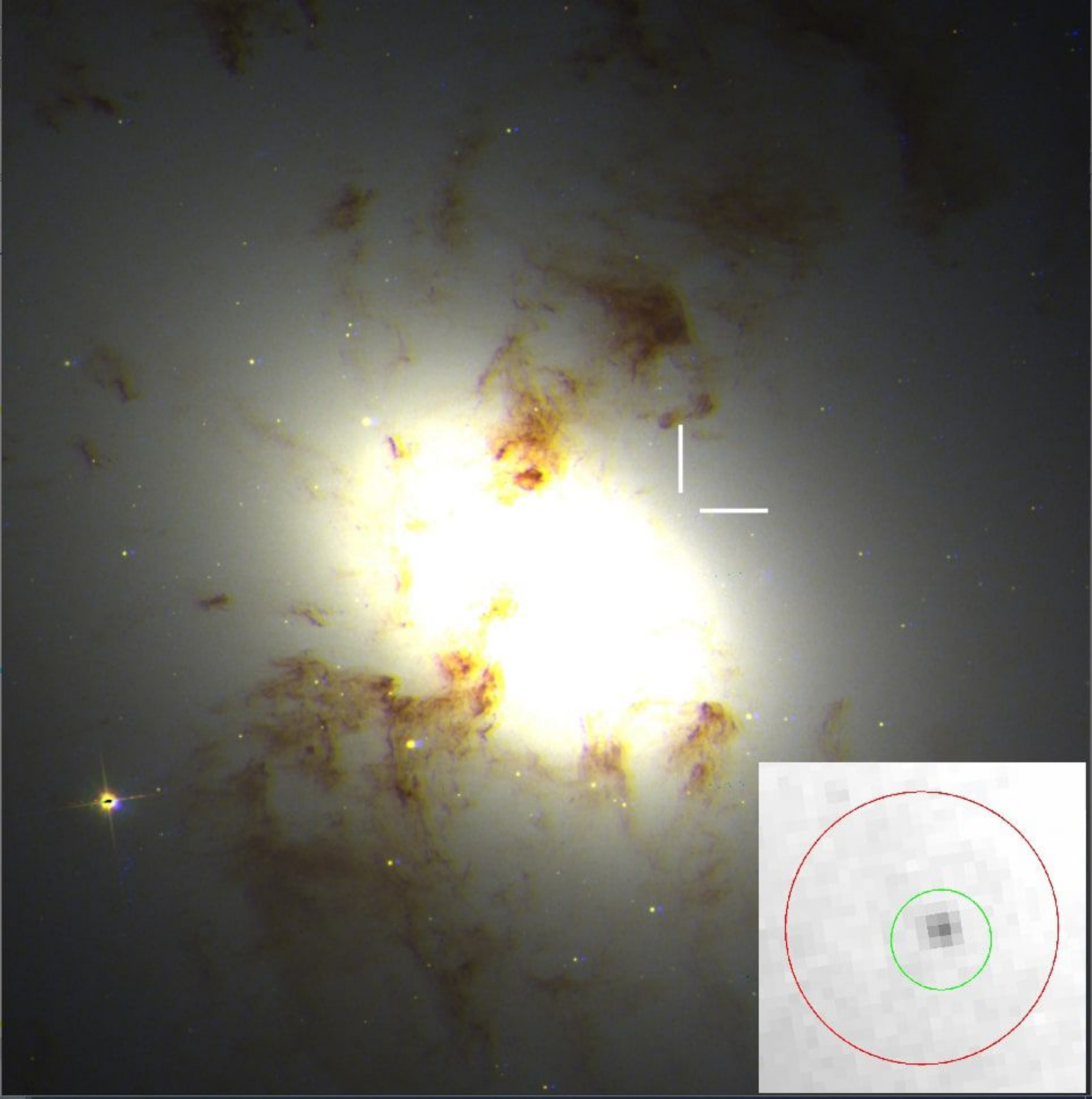}
\caption{Left panel: {\it {\it Chandra}/ACIS-S} image of the galaxy NGC 1316. Circles with 2.5\arcsec radius show the location of X-7 (magenta), X-5 (black), transient source XT-1 (blue) and background (red), respectively.
Right panel: {\it HST} image of the X-7 with RGB (Red: ACS/F814W, Green: ACS/F555W and Blue: WFC3/F336W). The position of X-7 is marked with white bars.
An inset represents inverted color {\it HST}/ACS/F555W image of the X-7. The red and green circles indicate the {\it Chandra} position of X-7 with a radius of 0.\arcsec6 and the corrected position of X-7 with an error radius of 0.\arcsec22, respectively. Both panels show exactly the same region. X axis is  R.A. and Y axis is Dec. The north is up and the east is left.}
\label{F:1}
\end{center}
\end{figure*}

{\it Chandra} data were analyzed by using {\scshape ciao}\footnote{https://cxc.cfa.harvard.edu/ciao/} v4.12 software with its calibration package {\scshape caldb}\footnote{https://cxc.cfa.harvard.edu/caldb/} v4.9. Several X-ray sources were detected from level 2 event list using {\scshape wavdetect} tool in {\scshape ciao}. We determined new coordinates for X-7 as R.A.$=$ 03$^{\mathrm{h}}$22$^{\mathrm{m}}$40$^{\mathrm{s}}.813$ and Dec.$=$ -37$^{\circ}$12$\arcmin$23$\arcsec.47$ by using C6 data. The differences between the early coordinates given by \cite{2003ApJ...586..826K} were $\sim$ 0$\arcsec$.20 for R.A and 0$\arcsec$.53 for Dec. The new ULX source X-7 is located 12$\arcsec$ away from center of the galaxy. The position of this source is shown on the {\it Chandra} and {\it HST} images in Fig. \ref{F:1}.

The source and background photons were extracted with {\scshape specextract} task using a circle with a radius of 2.5 arcsec. The background subtracted count rates and off-axis distances of the source X-7 are also given in Table \ref{T:1}. The off-axis distances were calculated with {\scshape src\_psffrac} task for 2$\arcsec$.5 radius in the 0.3-10 keV energy band. The spectral analyses of the X-ray data have been performed by using the package, {\scshape xspec} v12.11. We only used observations labeled as C5 (ObsID 20341) and C6 (ObsID 22187) for the spectral analyses due to the low statistics ($\leq$ 50 counts) of other {\it Chandra} 2019 observations.

Several single component models such as absorbed power-law ({\scshape pl}), multi-color disk blackbody ({\scshape diskbb}) and blackbody ({\scshape bbody}) were fitted to the spectra of X-7. We also fitted the source spectra with the frequently used two-component models such as {\scshape diskbb} + {\scshape pl} and {\scshape diskbb} + {\scshape comptt}; however, no statistically significant improvement was achieved for C5 and C6 data. Hence, we will not discuss the two component models any further in this work.

In our initial fits, each model included two absorption components: The first one was a fixed Galactic column density of N$_{H}$ = 2$ \times$ $10^{20}$ cm$^{-2}$ \citep{1990ARA&A..28..215D} representing the absorbing line-of-sight column density along to the NGC 1316 (using the {\scshape tbabs} model) and the other one was left free to account for intrinsic absorption. The latter component was found to be negligible in the resulting fitting parameters. The unabsorbed flux was calculated in the 0.3$-$10 keV energy band using the convolution model {\scshape cflux} available in {\scshape xspec} and the corresponding luminosity value was calculated using the distance of 19 Mpc. Due to the relatively low source counts for C5 (88 net counts) and C6 (102 net counts) observations we used the Cash statistic\footnote{https://heasarc.gsfc.nasa.gov/xanadu/xspec/manual/XSappendixStatistics.html} \citep[or C-stat;][]{1979ApJ...228..939C} for spectral fitting and spectra were grouped at least of 5 counts per bin. We used background subtracted source spectra in our analysis. The resulting spectral parameters of single-component models for C5 and C6 data are given in Table \ref{T:2}.

In order to search for any periodicity, C6 data were used to perform a timing analysis since it provides better statistics than others. The X-ray light curve of X-7 was sampled at 3.2 s using {\scshape dmextract} tool in {\scshape ciao} and power density spectra (PDS) were calculated using {\scshape xronos} v6.0 in {\scshape heasoft} v6.27. The PDS were calculated from a single interval or up to six spectra were averaged to produce a PDS in the 0.3$-$10 keV band. We cannot confirm any significant ($\geq$ 3$\sigma$) periodicity with time-bin size of 3.2 s which sets a lower limit for searching periods. We calculated the pulse fractions of possible sinusoidal modulations in the frequency range (0.001$-$0.1) Hz, following the approach given by \cite{2015MNRAS.452.1112E}. We inferred a 3$\sigma$ upper limit on the highest pulsed fraction as $\sim$ 26\%.

We also searched for short-term and long-term count rate variability. For the short-term variability, the light curve of X-7 was binned over intervals of 100s, 500s and 1000s in the 0.3$-$10 keV energy band using C6 data. The resulting light curves were tested for short-term variations in the source count rates using a Kolmogorov–Smirnov (K-S) test. The K-S test probabilities are found as > 0.3. These results indicate that X-7 do not show any significant amplitude variations.

Bi-modal distribution obtained from the long-term light curves is considered to be a good indication of the presence of a neutron star in ULXs \citep{2018MNRAS.476.4272E,2020MNRAS.491.1260S,2020ApJ...890..166P,2020ApJ...895..127B}. Therefore, we also checked the long-term variability of X-7, for this we used equation of \cite{2020ApJ...895..127B}. According to the their definition the count rate variability is determined the equality by $\chi_{r}^{2}$ $\equiv$ $\chi^{2}$\big/ N$_{obs}$, where $\chi_{r}^{2}$ is an arbitrary definition of count rate variability, $\chi^{2}$ is defined as ${\sum_{\substack{n={1}\\}}^{N_{obs}} \left( \frac{CR_{n}-<CR>}{\sigma_{n}} \right)^2}$, and N$_{obs}$ is the number of observations.
Here, CR$_{n}$ is the count rate in each observation, <CR> is the mean count rate averaged over all observations and $\sigma_{n}$ is the 1$\sigma$ uncertainty on the count rate for each observation. In their work, when $\chi_{r}^{2}$ > 2 the source is considered to be variable. They applied this formula to their sources in M51 and obtained $\chi_{r}^{2}$ is about 10 for ULX X-7 and ULX X-8. Then, they considered these two sources strongly variable. We found that $\chi_{r}^{2}$ $\simeq$ 5 for all {\it Chandra} observations (including 2001 observation) and comparing our value with theirs, X-7 of NGC 1316 is also probably mildly variable (within the present data coverage). Due to the fact that, more observations are required to interpret the long-term variability of X-7 might point out a neutron star in ULX system.

We also examined if luminosities of the ULXs (X-3, X-5 and X-4) in NGC 1316 given by \cite{2011ApJS..192...10L}, vary or not. We performed spectral analyses of these ULXs by using C6 data. The unabsorbed X-ray luminosity L$_{X}$ was calculated with an {\scshape pl} model in 0.3-8.0 keV adopted distance 19 Mpc for each of three sources. Resulting L$_{X}$ values are 4.11, 2.52, and 4.01$\times$ 10$^{39}$ erg s$^{-1}$ for X-3, X-5 and X-4, respectively. Variation of the L$_{X}$ values for these ULXs were found less than a factor two on a timescale of 18 years.

\begin{figure*}
\begin{center}
\includegraphics[angle=0, scale=0.60]{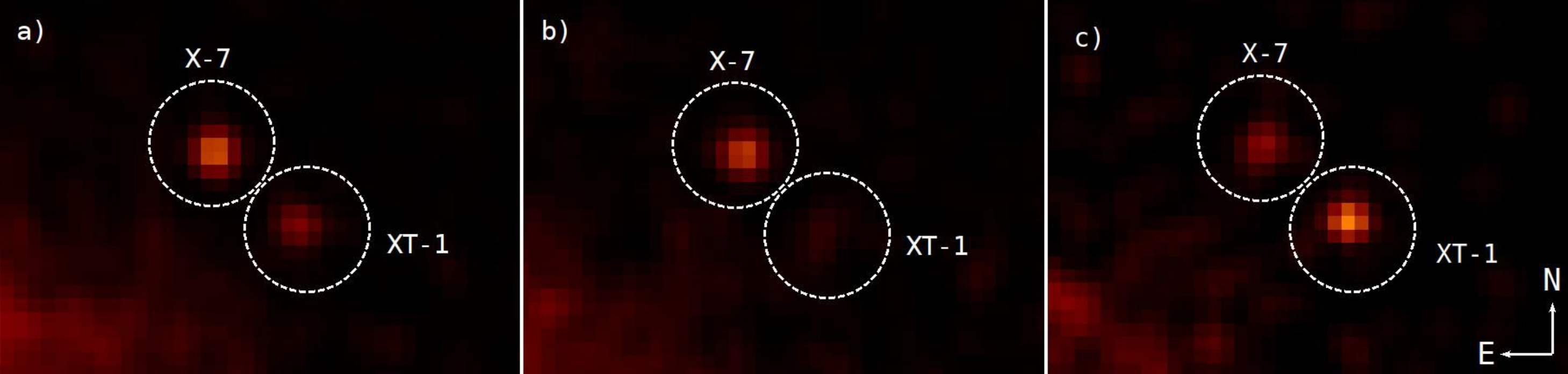}
\caption{Comparison of X-7 and XT-1 in the {\it Chandra} observations (C1:a, C2:b and C4:c). The X-7 and XT-1 positions are represented with dashed white circles. The XT-1 lies at $\sim$ 6$\arcsec$ to X-7. It is clearly seen that XT-1 is not detected in the C2 observation. The same regions of  19$\arcsec$ $\times$ 16$\arcsec$ are given in all panels.}
\label{F:4}
\end{center}
\end{figure*}

In addition, we determined a source located at $\sim$ 6$\arcsec$ southwest of the X-7 at R.A$=$ 3$^{\mathrm{h}}$22$^{\mathrm{m}}$40$^{\mathrm{s}}.550$ and Dec.$=$ -37$^{\circ}$12$\arcmin$26$\arcsec.45$ (see Fig. \ref{F:1}). This source is also in the list of \cite{2003ApJ...586..826K} as CXOU J032240.5-371227 (\#16, within the D$_{25}$ ellipse of NGC 1316) with a luminosity of 3.34$\times$10$^{38}$ erg s$^{-1}$.
The source count rates for all data (C1-C6) were obtained fitting an absorbed {\scshape pl} model with same N$_{H}$ $=$ 2$\times$ $10^{20}$ cm$^{-2}$ the $\Gamma$ $=$ 1.7 and resulting unabsorbed fluxes were obtained by using {\scshape srcflux} tool in {\scshape ciao} at 90\% confidence level in the range of 0.3$-$10 keV. 
The flux values of C2 and C4 data were found as (0.16$\pm$0.04$)\times$10$^{-14}$ erg cm$^{-2}$ s$^{-1}$ (minimum) and (2.31$\pm$0.56)$\times$10$^{-14}$ erg cm$^{-2}$ s$^{-1}$ (maximum). This maximum flux corresponds to luminosity of $\sim$1.0$\times$10$^{39}$ erg s$^{-1}$ with an adopted distance of 19 Mpc. We note that, the flux variation of the source is a quite significant with an order of magnitude difference. Due to the variation, this source could be a candidate of transient ULX which we named as XT-1, as seen clearly in Fig. \ref{F:4}.

\subsection{HST} 

The archival data log of Hubble Space Telescope\footnote{https://archive.stsci.edu/hst/search.php} ({\it HST}) observations used for analysis are given in Table \ref{T:1}. We performed a relative astrometry between {\it Chandra} (ObsID 22187) and the {\it HST}/ACS (Advanced Camera for Surveys) (ObsID j6n202010) images to find the optical candidate of X-7. We used {\it wavedetect} tool in {\scshape ciao} and {\it daofind} tool in {\scshape iraf} for source detection in {\it Chandra} and {\it HST}/ACS images, respectively.

Nine reference sources were determined from the comparison of the {\it Chandra} and {\it HST} images. All of these matched sources are located on ACIS-S with a moderate offset from the optical axis in the ObsID 22187. Properties of reference sources are summarized in Table \ref{T:3}. The position uncertainties between each of reference sources were calculated with 90\% confidence level. The final astrometric errors between the {\it Chandra} and {\it HST} images are in R.A. 0\farcs09 and in Dec. 0\farcs07. As a result of astrometric correction, the {\it HST} coordinates of X-7 are given as R.A$=$ 03$^{\mathrm{h}}$22$^{\mathrm{m}}$40$^{\mathrm{s}}.806$ and Dec.$=$ 37$^{\circ}$12$\arcmin$23$\arcsec.52$ within 95 \% confidence level of error circle with 0\farcs22 radius. 

The Point-Spread Function (PSF) photometry was performed with {\scshape dolphot} v2.0 \citep{2000PASP..112.1383D} using the {\it HST} data listed in Table \ref{T:1}. The {\scshape acsmask} task was used to remove pixels flagged as bad in images. {\scshape splitgroups} and {\scshape calcsky} tasks were performed to split into single-chip images and were created the sky background for each image, respectively. Here, we present the magnitudes derived by using the set of parameters for the ACS, WFC3/UVIS and WFC3/IR recommended by {\scshape dolphot} user’s guide. The {\scshape dolphot} task was used for photometry by taking drizzled images ObsID j6n202010 and ObsID ib3n03030 as the positional reference for both ACS and WFC3, respectively. 

We found a unique optical candidate within 0.\arcsec22 error circle at 95\% confidence level for X-7 using the archival {\it HST}/ACS and {\it HST}/WFC3 data (see \ref{F:1}). The ACS and WFC3 magnitudes in the VegaMag. and Johnson system for the optical candidate are given in Table \ref{T:4}. Magnitude values were corrected with $E$($B-V$) = 0.018 mag. derived from the Galactic extinction (A$_{V}$/3.1). The A$_{V}$ $=$ 0.058 mag. was obtained from extinction calculator tools of NED \footnote{https://ned.ipac.caltech.edu/extinction\_calculator}. The F475W, F555W and F814W filters in {\it HST}/ACS correspond to the Johnson filter {\it B}, {\it V} and {\it I}, respectively \citep{2005PASP..117.1049S, 2011PASP..123..481S}. We obtained {\it B$-$V}$=$0.77 mag., {\it V$-$I}$=$1.10 mag. and M$_{V}$$=$$-$7.80 mag. The distance modulus of NGC 1316 is calculated as 31.4 mag. by using the adopted distance of 19 Mpc.

\begin{figure*}
\begin{center}
\includegraphics[angle=0, scale=0.60]{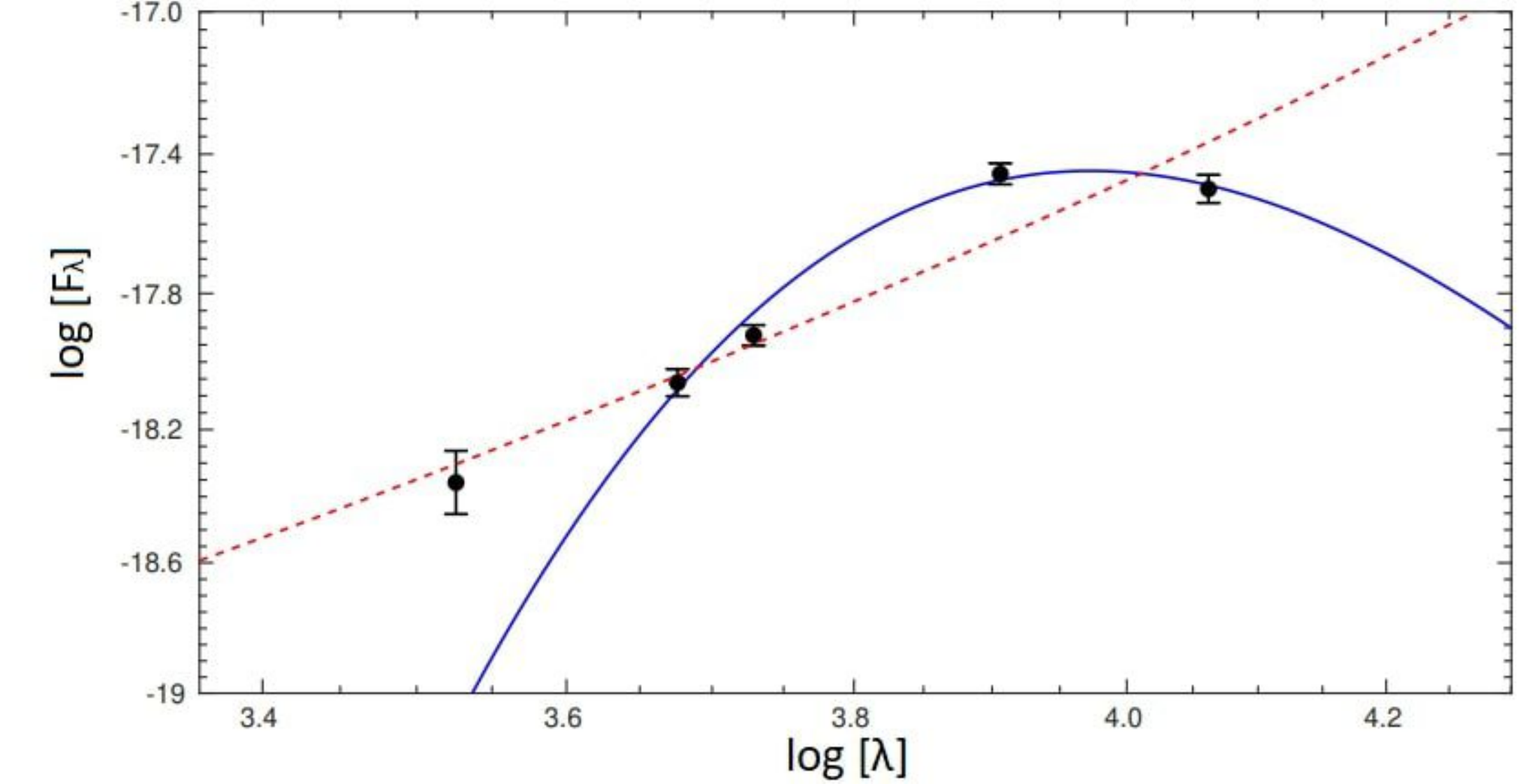}
\caption{The SED of optical candidate of X-7. All data are shown with black circles and their respective error with bars. The SED models are shown by blue solid line for blackbody and red dashed line for power-law. The blackbody has a temperature of 3100 $\pm$ 400 K and the power-law model is $\alpha$ = 1.75 $\pm$ 0.35 for the X-7. The units of y and x axes are erg s$^{-1}$ cm$^{-2}$ and \AA, respectively.} 
\label{F:bb}
\end{center}
\end{figure*}

Spectral Energy Distribution (SED) of the optical candidate has been constructed to obtain the spectral characteristics of X-7 using the derived flux values given in Table \ref{T:1}. The wavelength of the filters are selected as the pivot wavelength, obtained from {\scshape pysynphot}\footnote{https://pysynphot.readthedocs.io/en/latest/}, in SED plots. The SED for the optical candidate is adequately fitted (1) a blackbody spectrum with a temperature of 3100 $\pm$ 400 K or (2) a power-law spectrum (F $\propto$ $\lambda^{\alpha}$) with $\alpha$ = 1.75 $\pm$ 0.35, see Fig. \ref{F:bb}. To obtain a blackbody spectrum, a code has been used with {\scshape optimset} and {\scshape fminsearch} functions in {\scshape matlab}\footnote{https://www.mathworks.com/matlabcentral/fileexchange/20129-fit-blackbody-equation-to-spectrum}. The reduced $\chi^{2}$ for blackbody and power-law are 1.34 and 1.27, respectively. The number of degrees of freedom is 3 for both models. 

\section{Results and discussions} \label{section:3}

\subsection{X-Rays} 

\begin{figure*}
\begin{center}
\includegraphics[angle=-90, scale=0.05]{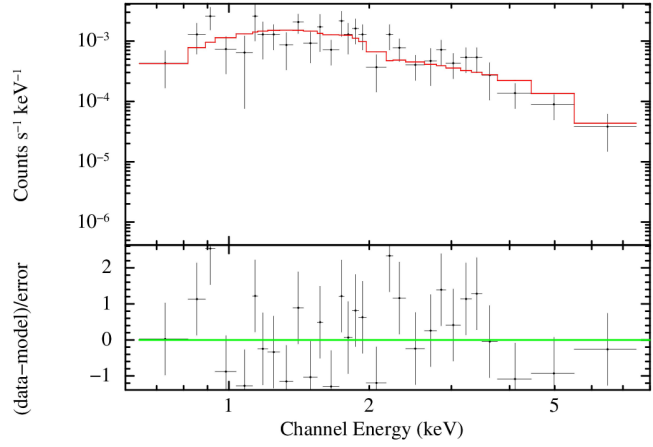}
\caption{Best-fitting model spectrum (upper panel) and the residual between the data and the model  (lower panel) of the X-7 are shown for C6 data. The spectrum was fitted with a {\scshape pl} model.
}
\label{F:3}
\end{center}
\end{figure*}

We identified a new ULX in NGC 1316 by {\it Chandra} in April 2019 with at a luminosity of 2.10 $\times$ 10$^{39}$ erg s$^{-1}$. This luminosity is almost a factor of $\sim$ 4 times higher than previous {\it Chandra} observation in April 2001. We obtained the best-fitting spectral parameters of one-component models for X-7 to elaborate its spectral characteristics. These parameters were obtained with only one intrinsic absorption, N$_{H}$ $=$ 2 $\times$ $10^{20}$ cm$^{-2}$ since the second absorption parameter did not contribute significantly. 
The spectra of X-7 are the best fitted by {\scshape pl} models with photon indices of $\Gamma$ $=$ 2.24 for C5 and 1.56 for C6 data. The source has maximum unabsorbed luminosity values of $\sim$ 1.8 and 2.1$\times$ $10^{39}$ in 0.3$-$10 keV, respectively. These values are also listed in Table \ref{T:2}. Energy spectrum of X-7 with a {\scshape pl} model and its residual are shown in Fig. \ref{F:3}.
The source spectra were also fitted by the {\scshape diskbb} model with T$_{in}$ $=$ 0.78 keV (C5) and 1.32 keV (C6). The other well fitted model is {\scshape bbody} whose temperature values are 0.50 keV and 0.74 keV with unabsorbed luminosities 0.92 and 1.21 $\times$ $10^{39}$ erg s$^{-1}$ for C5 and C6, respectively.
As noted, it is difficult to differentiate between these three models due to their very similar C-stat.


In the spectral analyses of the {\it Chandra} data, we found that the spectra of X-7 in C5 and C6 are better represented by {\scshape pl} models with photon indices $\Gamma$ $\sim$ (2.24 and 1.56). Both of these $\Gamma$ values correspond to hard states defined for GBHBs. Hard states with low luminosity are seen at sub-Eddington mass accretion rates. On the other hand, {\scshape diskbb} model also yields acceptable fits for the same datasets with the temperature of $kT_{in}$ $\sim$ (0.78 and 1.32) keV. These $kT_{in}$ values are compatible with those of GBHBs at a high mass accretion rate during thermal state \citep{2006ARA&A..44...49R}. Generally in GBHBs, luminosities are usually hard when they are bright and soft when they are dim. However, X-7 exhibits the opposite behavior since the source has a high $L_{X}$ when it is in a hard state and a low $L_{X}$ when it is in a soft state. There are some ULXs that do show similar correlations like X-2 of NGC 1313, \cite{2006ApJ...650L..75F}; X-1 of IC 342, \cite{2014MNRAS.444..642M}; X-2 of NGC 4736, \citet{2014Ap&SS.352..123A}. X-3 of NGC 925, \cite{2020ApJ...891..153E}.

We used the normalization parameter of the {\scshape diskbb} model in C6 data to constrain the mass of the compact object. Here, we calculated the inner disk radius as R$_{in}$ $\sim$ 70 km from the equation of R$_{in}$ = $\kappa$ $^{2}$ $\xi $r$_{in}$ $\sqrt{cos\theta}$, where $\xi$ = 0.412 is the correction factor, $\kappa$ = 1.7 is spectral hardening factor, r$_{in}$ is the apparent disk radius derived from the observed data and $\theta$ is the disk inclination angle and its range was given as 0$^{\circ}$-75$^{\circ}$. \citep{1995ApJ...445..780S, 1998PASJ...50..667K} Using the relation between inner disk radius and mass \citep{2000ApJ...535..632M}, we found the average value $\sim$ 8 M$\odot$ for the mass of compact object in X-7.

\subsection{HST}
 
We examined the optical properties of X-7 in the galaxy NGC 1316 using the archival data from {\it HST}/ACS and {\it HST}/WFC3. A unique optical candidate was identified within $0.\arcsec22$ error radius after the astrometric correction. It is noted that dereddened magnitudes of optical candidate are faint (m$_{V}$ > 22.5 mag) as seen in Table \ref{T:4}. On the other hand, the absolute magnitude is very bright M$_{V}$ $=$ $-$ 7.80 mag which is the compatible with the given range ($-$4 < M$_{V}$ < $-$9 ) for ULXs \citep{2013ApJS..206...14G}.

Assuming optical emission originates from the donor star, we fitted a blackbody and power-law model to SED of X-7 as seen in Fig \ref{F:bb}. The best-fitting parameter for power-law model is $\alpha$ $=$ 1.75 $\pm$ 0.35 and the temperature is 3100 $\pm$ 400 K for blackbody model. Although, both model yields similar reduced $\chi^{2}$ $\simeq$ 1.3, the $\alpha$ parameter for power-law is not in the range of $-$3 and $-$4. This range is given for optical counterparts of many ULXs having power-law SEDs \citep{2011ApJ...737...81T, 2018ApJ...854..176V}. It follows that the blackbody model is more favorable in our case. According to \cite{1981Ap&SS..80..353S} the measured temperature and luminosity of the companion star indicate that spectral classification would be an M type supergiant. Similar spectral type of companions are proposed for ULX-5 of NGC 3034, ULX-1 and ULX-2 of NGC 253 by \cite{2013ApJS..206...14G}.

The blackbody emission is also thought to be originating from irradiation of the accretion disk. Assuming so and taking into consideration the temperature from blackbody fitting and the distance to the source, we determined the surface area of the emitting region of X-7 as $\simeq$ 4 $\times$ 10$^{26}$ cm$^{2}$ or a radius of $\simeq$ 2 $\times$ 10$^{13}$cm. This radius seems an order of magnitude larger than those of the optical emitting region estimated ULXs \citep{2011NewAR..55..166F}. \cite{2011ApJ...737...81T} also fitted these two models to the spectrum of X-1 in IC 342 and showed that their data is better fitted to the blackbody model and reported very similar radius. 

We have also examined whether the size of the emitting region can reprocess this optical emission. Assuming a fraction $\eta$ of the X-ray luminosity, L$_{X}$, is reprocessed with $\eta$ $=$ $\sigma$T$^{4}$A/L$_{X}$ where $\sigma$, T and A are Stefan–Boltzmann constant, temperature and area of the region, respectively. Inserting T$=$3100 K, A $\sim$ 10$^{27}$ cm$^{2}$, and L$_{X}$ $=$ 2$\times$10$^{39}$ erg cm$^{-2}$ s$^{-1}$, we found that $\eta$ is $\simeq$ 0.002 which is similar to the value for GBHBs such as XTE J1817$-$330 and X-1 of Holmberg IX \citep{2009MNRAS.392.1106G,2011ApJ...737...81T}.

In addition, we investigated the optical candidates of previously known ULXs (X-3, X-4 and X-5) from study of \cite{2011ApJS..192...10L} and also the XT-1 in the NGC 1316. As noted, we identified a single optical candidate for X-5 located at the edge of 0\arcsec.22 error circle. Its V-band magnitude is 23.1$\pm$0.03 mag. This candidate is not associated with any source in the CDS (Centre de Données astronomiques de Strasbourg) database. On the other hand, the source XT-1 has no optical counterpart(s) in {\it HST} images.\\

As a conclusion, further broadband X-ray observations of X-7 are needed to constrain the spectral parameters better and scale of variability of the source with more confidence. It will allow us to interpret mechanisms of X-ray emission from accretion processes in ULXs. 

\section*{Acknowledgements}
This research was supported by the Scientific and Technological Research Council of Turkey (TÜBİTAK) through project number 117F115. This research was also supported by the Çukurova University Research Fund through project number FBA-2019-11803. This research is a part of the PhD thesis of S. Allak and he acknowledges financial support by TÜBİTAK. Our special thanks to M. E. Ozel for his valuable contributions of the manuscript. 

\section*{Data Availability}
The scientific results reported in this article are based on archival observations made by the Chandra X-ray Observatory\footnote{https://cda.harvard.edu/chaser/}. This work has also made use of observations made with the NASA/ESA Hubble Space Telescope, and obtained from the data archive at the Space Telescope Science Institute\footnote{https://mast.stsci.edu/portal/Mashup/Clients/Mast/Portal.html}.




\bibliographystyle{mnras}
\bibliography{ngc1316} 


\begin{table*}
\centering
\caption{The log of X-ray and optical observations of X-7}
\begin{tabular}{ccccccc}
\hline\hline
Instruments & Label/Filter & ObsID & Date & Exp. & Rate$^{a}$ & Off-axis$^{b}$  \\
\hline
--&--&--&(YYYY-MM-DD)&(ks)&10$^{-4}$(count/s)&($\arcmin$)\\
\hline
{\it Chandra}/ACIS-S  & C1     & 2022      & 2001-04-17 & 29.85 & 17.01  $\pm$ 3.48 & 1.14\\
{\it Chandra}/ACIS-S  & C2     & 20340     & 2019-04-16 & 44.97 & 7.86   $\pm$ 1.74 & 1.07\\
{\it Chandra}/ACIS-S  & C3     & 22179     & 2019-04-17 & 38.95 & 10.80  $\pm$ 2.04 & 1.07\\
{\it Chandra}/ACIS-S  & C4     & 22180     & 2019-04-20 & 13.57 & 6.65   $\pm$ 3.29 & 1.13\\
{\it Chandra}/ACIS-S  & C5     & 20341     & 2019-04-22 & 51.39 & 17.20  $\pm$ 2.24 & 1.12\\
{\it Chandra}/ACIS-S  & C6     & 22187     & 2019-04-25 & 53.18 & 19.41  $\pm$ 2.30 & 1.10\\
\hline
{\it HST}/ACS/WFC  &  F555W & j6n202010  & 2003-03-04  & 6.98  &-- &--\\	
{\it HST}/ACS/WFC  &  F814W & j6n201030  & 2003-03-07  & 2.20   &--&--\\
{\it HST}/ACS/WFC  &  F475W & j90x01020  & 2005-02-16  & 0.76   &--&--\\
{\it HST}/WFC3/UVIS &  F336W & ib3n03040  & 2010-07-30  & 3.23  &--&--\\
{\it HST}/WFC3/IR &  F110W & ib3n03030  & 2010-07-30  & 0.40   &--&--\\ 
\hline
\end{tabular}
\label{T:1}
\\ $^{a}$ The background subtracted count rates were calculated in the 0.3-10 keV energy band by using {\scshape xspec}. $^{b}$ Off-axis distances of the source X-7.
\\
\end{table*}

\begin{table*}
\centering
\caption{X-ray spectral fitting parameters of X-7.}
\begin{tabular}{ccccccccccc}
\hline\hline
Data & Model Name & $\Gamma$ & T$_{in}$$_{(DISKBB)}$/kT$_{(BBODY)}$ & N$_{\mathrm{(PL)}}$$^{\textcolor{blue}{1}}$ & N$_{\mathrm{(DISKBB)}}$$^{\textcolor{blue}{2}}$ & N$_{\mathrm{(BBODY)}}$$^{\textcolor{blue}{3}}$ & C/dof$^{\textcolor{blue}{4}}$ & L$_{\mathrm{X}}$$^{\textcolor{blue}{5}}$ \\
& -- & -- & (keV) & (10$^{-6}$) & (10$^{-4}$) & (10$^{-7}$) &--& (10$^{39}$ erg s$^{-1}$)\\
\hline
\multirow{3}{*}{C5} & {\scshape pl} & $2.24_{-0.36}^{+0.37}$ &--& $8.07_{-1.82}^{+2.13}$ & -- &  & 23.79/21(1.13) & $1.77_{-0.23}^{+0.27}$ \\
& {\scshape diskbb} &--& $0.78_{-0.06}^{+0.07}$ &--&  $0.31_{-0.05}^{+0.05}$ & -- & 21.11/21(1.01)& $1.10_{-0.13}^{+0.17}$  \\	
&{\scshape bbody}&--& $0.50_{-0.06}^{+0.07}$ & -- & -- & $2.57_{-0.40}^{+0.45}$ & 25.80/21(1.23) & $0.92_{-0.07}^{+0.11}$ \\
\hline
\multirow{3}{*}{C6} &{\scshape pl} & $1.56_{-0.27}^{+0.29}$ &--& $5.20_{-1.30}^{+1.42}$ & -- & -- & 27.58/27(1.02) & $2.10_{-0.26}^{+0.29}$ \\
 & {\scshape diskbb} &--& $1.32_{-0.27}^{+0.51}$ &--& $4.80_{-0.76}^{+0.76}$ & -- & 24.71/27(0.92)& $1.47_{-0.20}^{+0.23}$  \\	
 &{\scshape bbody}&--& $0.74_{-0.10}^{+0.11}$ & -- & -- & $3.35_{-0.58}^{+0.68}$ & 26.14/27(0.96) & $1.21_{-0.19}^{+0.23}$ \\
\hline     
\end{tabular}
\label{T:2}
\\\textbf{Notes}: A fixed N$_{H}$=2x$10^{20}$ was used in all models. \\$^{\textcolor{blue}{1}}$ {\scshape pl} model normalization parameter. 
$^{\textcolor{blue}{2}}$ {\scshape diskbb} model normalization parameter;  N = (r$_{in}$ / D$_{10}$)$^{2}$ $\times$ cos$\theta$, where r$_{in}$ is the apparent inner disk radius, D is the distance to the source in units of 10 kpc, and $\theta$ is the inclination of the disk. 
$^{\textcolor{blue}{3}}$ {\scshape bbody} model normalization parameter. 
$^{\textcolor{blue}{4}}$ C-statistic of the fit and the number of degrees of freedom. $^{\textcolor{blue}{5}}$ The unabsorbed luminosity were calculated in energy range of 0.3-10 keV at a distance of 19 Mpc. \\
\end{table*}

\begin{table*}
\centering
\caption{Coordinates of the X-ray/optical reference sources. {\it Chandra} ACIS-S sources (ObsID 22187) identified with {\it HST} data (j6n202010).}
\begin{tabular}{ccccccccccc}
\hline\hline
{\it Chandra} R.A.& {\it Chandra} Dec.& Net Counts$^{a}$
 & {\it HST} R.A.& {\it HST} Dec.& Position Uncertainty$^{b}$ \\
(hh:mm:ss.sss) & ($\degr$ : $\arcmin$ : $\arcsec$) & -- & (hh:mm:ss.sss) & ($\degr$ : $\arcmin$ : $\arcsec$) & ($\arcsec$)\\
\hline
3:22:42.719 &-37:12:16.63 & 39.20 $\pm$7.00 & 3:22:42.728 &-37:12:16.60 & 0.16\\
3:22:42.066 &-37:12:18.23 & 36.11 $\pm$ 7.14& 3:22:42.055 &-37:12:18.26 & 0.13\\
3:22:42.484 &-37:12:21.06 & 56.50 $\pm$ 9.11 & 3:22:42.457 &-37:12:20.99 & 0.37\\
3:22:42.596 &-37:12:06.32 & 08.92 $\pm$ 3.16  & 3:22:42.600 &-37:12:06.24 & 0.11\\
3:22:41.153 &-37:11:43.92 & 15.70 $\pm$ 4.58 & 3:22:41.172 &-37:11:44.26 & 0.48\\
3:22:41.318 &-37:11:17.09 & 22.02 $\pm$ 5.00 & 3:22:41.297 &-37:11:17.13 & 0.29\\
3:22:39.123 &-37:11:47.97 & 66.18 $\pm$ 8.49 & 3:22:39.120 &-37:11:48.00 & 0.05\\
3:22:37.710 &-37:12:51.05 & 18.11 $\pm$ 4.47 & 3:22:37.690 &-37:12:51.13 & 0.28\\
3:22:44.729 &-37:13:10.03 & 09.76 $\pm$ 3.32 & 3:22:44.713 &-37:13:10.09 & 0.22\\
\hline
\end{tabular}\\
\label{T:3}
$^{a}$The counts were calculated in the 0.3-10 keV using {\scshape xspec}.\\
$^{b}$Uncertainties are given at 90\% confidence level of the {\it Chandra}/{\it HST} reference sources.
\end{table*}

\begin{table*}
\centering
\caption{The dereddened {\it HST} magnitudes and color values of optical candidate of X-7.}
\begin{tabular}{lcccccc}
\hline\hline
Instruments & Date & Filter & Pivot Wavelength$^{\clubsuit}$& VegaMag & JohnsonMag  & Flux \\
&(yyyy-mm-dd)&&({\AA})&&&10$^{-18}$ (erg cm$^{-2}$ s$^{-1}$ Å$^{-1}$)\\
\hline
\multirow{3}{*}{ACS} &2005-02-16 & F475W & 4746.7 & $24.05\pm 0.06$ & (B) $24.40\pm 0.06$  & $0.86\pm0.02$ \\
&2003-03-04 & F555W & 5360.8 & $23.70\pm 0.03$& (V) $23.63\pm 0.03$  & $1.19\pm 0.01$\\
&2003-03-07 & F814W & 8044.8 & $22.54\pm 0.02$& (I) $22.53\pm 0.02$  & $3.49\pm 0.01$\\ 
&&&&& $(B-V)_{0}$ = 0.77 &   \\ 
&&&&& $(V-I)_{0}$ = 1.10 &   \\ 
&&&&&M$_{V}$ = -7.80 &       \\
\multirow{2}{*}{WFC3}&2010-07-30 & F336W & 3354.8  & $24.79\pm 0.21$  &--& $0.43\pm 0.08$ \\
&2010-07-30 & F110W & 11534.5 & $22.64\pm 0.15$  &--& $3.16\pm 0.02$\\
\hline
\end{tabular}\\
\label{T:4}
$^{\clubsuit}$ The Pivot Wavelengths were calculated using {\it pysynphot}.\\
\end{table*}

\bsp	
\label{lastpage}

\end{document}